\documentclass[useAMS,usenatbib,usegraphicx]{mn2e}

\usepackage{upgreek}
\usepackage{amsmath}
\usepackage{txfonts}
\usepackage{booktabs}
\usepackage{color}
\usepackage{array}
\usepackage{graphicx}
\usepackage{epstopdf}
\usepackage{bm}
\usepackage{cmap}
\usepackage{tabularx}
\usepackage{multicol}
\usepackage{natbib}
\usepackage{float}
\usepackage{multirow}
\usepackage{ulem}

\newcommand{\XMMU}{XMMU~J1732}
\newcommand{\fl}{f_{\ell  \rm p}}
\newcommand{\fln}{f_{\ell  \rm n}}
\newcommand{\Ts}{T_{\rm s}^\infty}
\newcommand{\DMC}{\Delta M_\mathrm{C}}
\newcommand{\Tc}{T_\mathrm{cp}}
\newcommand{\rhoC}{\rho_\mathrm{C}}
\newcommand{\gcc}{\mathrm{g~cm}^{-3}}
\newcommand{\Tg}{\widetilde{T}}
\newcommand{\xr}{x_\mathrm{r}}

\begin{document}

\title[Neutron Star in HESS J1731--347]{
Analyzing Neutron Star in HESS J1731--347 from Thermal Emission and
Cooling Theory}

\author[D. D. Ofengeim et al.]{
D. D. Ofengeim$^{1,2}$\thanks{E-mail: ddofengeim@gmail.com},
A. D. Kaminker$^2$,
D.  Klochkov$^3$,
V.  Suleimanov$^{3,4}$,
\newauthor{D. G. Yakovlev$^{2}$}\\
$^{1}$ St.~Petersburg Academic University, 8/3 Khlopina St.,
St.~Petersburg 194021, Russia\\
$^{2}$ Ioffe Physical Technical Institute, 26 Politekhnicheskaya
St., St.~Petersburg 194021, Russia\\
$^{3}$ Institut f{\"u}r Astronomie und Astrophysik, Universit{\"a}t
T{\"u}bingen (IAAT), Sand 1, 72076 T{\"u}bingen, Germany\\
$^{4}$Kazan (Volga region) Federal University, Kremlevskaya 18, 420008 Kazan, Russia}

\date{Accepted . Received ; in original form}
\pagerange{\pageref{firstpage}--\pageref{lastpage}} \pubyear{2015}

\maketitle


\begin{abstract}
The central compact object in the supernova remnant HESS J1731--347
appears to be the hottest observed isolated cooling neutron star.
The cooling theory of neutron stars enables one to explain observations
of this star by assuming the presence of strong proton superfluidity
in the stellar core and the existence of the surface heat blanketing
envelope which almost fully consists of carbon. The cooling model of
this star is elaborated to take proper account of the neutrino emission
due to neutron-neutron collisions which is not suppressed by proton
superfluidity. Using the results of spectral fits of observed
thermal spectra for the distance of 3.2 kpc and the cooling theory
for the neutron star of age 27 kyr, new constraints on the stellar
mass and radius are obtained which are more stringent than those
derived from the spectral fits alone.
\end{abstract}


\section{Introduction}
\label{sec:intro}

The neutron star XMMU J173203.3--34418 (hereafter \XMMU) belongs to
the class of central compact objects (CCOs), relatively young
cooling neutron stars in supernova remnants. CCOs possess relatively
low  dipole-like large-scale surface magnetic fields ($\lesssim
10^{10}-10^{12}$\,G) and show thermal emission in soft X-rays.
\XMMU\ was discovered in 2007 with {\it XMM-Newton}
\citep{Tianetal2010} near the center of  the HESS J1731--347 (=
G353.6--0.7) supernova remnant. It was observed also with {\it
Suzaku} and {\it Chandra} (e.g. \citealt{HG2010}). Spectral fits of
the X-ray emission from \XMMU\ with traditional black-body and
hydrogen atmosphere models yielded radius of the emitting region
much lower than typical radii of neutron stars for the plausible
distance to the source of 3--5\,kpc. These results might have been
interpreted as the radiation from a hot spot on the neutron star
surface. However, no pulsations have been detected from \XMMU\ so
far which would imply a very special geometry in which either the
spot is aligned with the spin axis of the star or the spin axis is
directed along the line of sight.

This uncomfortable explanation was questioned by
\citet{Klochkov2013}. The authors fitted the spectra with the carbon
atmosphere model and showed that such fits lead to the radius of the
emitting region comparable to a typical radius of neutron stars. If
so, the observed radiation can be interpreted as the thermal radiation
emergent from the entire neutron star surface. We adopt this
explanation here. We note, that \XMMU\ is the second CCO where the
carbon atmosphere is used to explain the observed emission
properties, after the neutron star in the Cassiopeia~A supernova
remnant \citep{HoHeinke_09}.

The next important step was done by \citet{Klochkov_etal15},
hereafter Paper I, who analyzed longer observations of \XMMU\ with
{\it XMM-Newton} and fitted the entire dataset with the carbon
atmosphere model computed by \citet{Suleimanov2014}.  They analyzed
different assumptions on the distance $d$  to the star and
concluded that $d=3.2$~kpc seems most
realistic (although other possibilities are not excluded).
Assuming $d=3.2$~kpc
they obtained sufficiently narrow confident regions for the neutron
star mass and radius.
As for the neutron star age $t$, they took $t=27$ kyr from
\citet{Tianetal2008} and added (rather arbitrarily) an uncertainty
range of 10--40 kyr as the age might in fact be lower. We adopt
these values here and apply the elaborated cooling model to analyze
the data.

It is well known that the cooling theory allows one to study
the properties of superdense matter in neutron star cores and constrain
the parameters of neutron stars (e.g., \citealt{YP04,Page_etal09}).
Here, we continue the theoretical interpretation of \XMMU\ with the
cooling theory started in Paper I but using a more elaborated
cooling model.

\renewcommand{\arraystretch}{1.10}
\begin{table}
\centering \caption{Four models of \XMMU\ used in fig.\ 7 of Paper
I and in Fig.\ \ref{f:cool}.
MU, nn, np, and pp indicate the neutrino cooling due to modified
Urca (MU) process, as well as due to three weaker reactions of
nucleon-nucleon collisions. The sixth column indicates the composition
of the heat blanketing envelope. The last column gives the theoretical
surface temperature $\Ts$ at $t=27$~kyr. Only the last (SFac) model
is consistent with observations. See text for details. }
    \begin{tabular}{l l  l l l l l}
    \toprule
    Model  &  MU  & nn & np & pp & Heat blanket & $T_{\rm s}^\infty$ [MK] \\
    \midrule
    MU  &  on &  on    &  on & on & iron & 0.96
 \\
    MUac  &  on  &  on     &  on & on & carbon & 1.23
\\
    SF   &  off  & on     &  off & off & iron & 1.34
\\
    SFac  & off  &  on     &  off & off & carbon & 1.77
\\
    \hline
  \end{tabular}
\label{tab:XMMU}
\end{table}
\setlength{\tabcolsep}{6pt}
\renewcommand{\arraystretch}{1.0}

Among cooling isolated middle-aged neutron stars whose thermal
surface radiation has been detected, \XMMU\ is a special case. At
$d=3.2$ kpc the effective surface temperature, as measured by a
distant observer, is $\Ts \sim 2$~MK (see Paper I and references
therein). If $t=27$ kyr,  \XMMU\ is the hottest {\bf
cooling} neutron star with the measured surface temperature.
 Of course, we do not mean magnetars which can be hotter because
of their magnetic activity.

\begin{figure}
\centering
\includegraphics[width=0.48\textwidth]{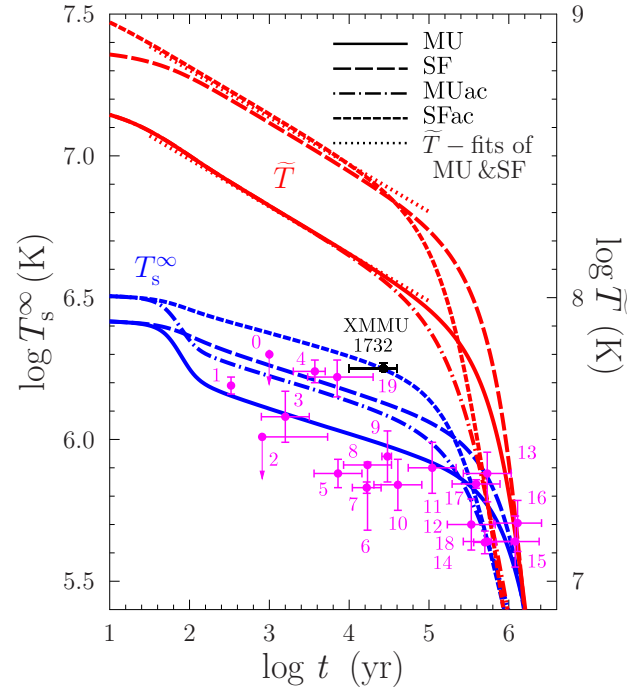} 
\caption{The effective surface temperatures $\Ts$ or upper limits
(left vertical scale) for a number of isolated neutron stars
including \XMMU\ versus their ages (data points; see the text). The data are
compared with the four theoretical cooling curves for a 1.5$\,{\rm
M_\odot}$ neutron star (Paper I). Curves MU and SF refer to neutron
stars without superfluidity and with strong proton superfluidity,
respectively, which have iron heat blanketing envelopes. Curves MUac
and SFac refer to similar neutron stars but with carbon envelopes.
The four upper curves show evolution of the internal temperature $\Tg$
(right vertical scale) for the same cooling scenarios. Two dotted
curves are analytic approximations of $\Tg$  described  in Section
\ref{s:coolfun}. } \label{f:cool}
\end{figure}

This special status of \XMMU\ is illustrated in Fig.\
\ref{f:cool} (after fig.\ 7 of Paper I)
which shows the measured temperatures  $\Ts$
(left vertical scale) and ages
$t$ of cooling neutron stars. The data and
neutron star labels are the same as in Paper I,
(0)~ the Crab pulsar; (1)~ the neutron star in Cas A; (2)~the
neutron star in 3C 58; (3)~PSR J1119--6127; (4)~RX J0822--43;
(5)~PSR J1357--6429; (6)~RX J0007.0+7303; (7)~the Vela pulsar;
(8)~PSR B1706--44; (9)~PSR J0538+2817; (10)~PSR B2334+61; (11)~PSR
0656+14; (12)~Geminga; (13)~PSR B1055--52; (14)~RX J1856--3754;
(15)~PSR J2043+2740; (16)~RX J0720.4--3125; (17)~RX J1741--2054;
(18)~PSR J0357+3205; (19)~1E 1207--52.

The data are compared with four theoretical cooling curves which
correspond to the four cooling models of \XMMU\
indicated in Table \ref{tab:XMMU} and detailed below. These
cooling curves have been calculated in Paper I for a neutron star
with the gravitational mass $M=1.5\,{\rm M_\odot}$ and
circumferential radius $R=12.04$ km (with a typical APR I equation of
state -- EOS -- in the stellar core, \citealt{GKYG05}). The four lower curves show the
evolution of the surface temperature $\Ts(t)$, while
the four upper
curves
show the evolution of the redshifted
temperature $\Tg(t)$ in the stellar centre.
The surface temperature inferred from observations
of \XMMU\ is $\Ts = 1.78^{+0.04}_{-0.02}$ MK (Paper I); it can be
compared with the theoretical values ($t=27$ kyr) presented
in Table \ref{tab:XMMU} for the four models.

The hot \XMMU\ can be explained by the standard cooling theory
although with great difficulty. It is thought to be an isolated
neutron star born hot in a supernova explosion. It gradually cools
down, and the cooling depends on properties of matter and neutron
star parameters. The cooling theory (e.g. \citealt{YP04}) states
that \XMMU\ is at the neutrino cooling stage with isothermal
interior; its internal thermal relaxation should be over. Such a
star should mainly cool via the neutrino emission from its super-dense
core (e.g. \citealt{YKGH01}). As in Paper I, we assume that the core
consists of nucleons, electrons and muons, and the nucleons can be
in superfluid state. It is well known (e.g., \citealt{LS01}) that
protons in the core are usually paired in singlet state while
neutrons can be paired in triplet state; singlet-state pairing of
neutrons can occur in the neutron star crust.

As the internal layers of \XMMU\ have to be isothermal, a
substantial temperature gradient is kept only in a thin (no thicker
than a few tens m) heat blanketing envelope. A  large-scale
surface magnetic field
$B\lesssim 10^{12}$ G which can exist in \XMMU\ (Paper I) cannot
greatly affect the cooling of this star. While analyzing the cooling
it can be neglected.

Because \XMMU\ is very hot, its cooling must be extremely slow.
Theoretically, such a cooling can be explained (Paper~I) by (i) a
strong suppression of neutrino emission from the stellar core and
(ii) assuming a massive carbon heat blanketing envelope. These two
cooling regulators are most important here.

(i) No enhanced neutrino cooling mechanism such as direct Urca
process \citep{LPPH91} or neutrino emission due to triplet-state
Cooper pairing of neutrons (\citealt{FRS1976,LP2006}; also see
\citealt{Page_etal09} and references therein) can operate in the
\XMMU\ core. The enhanced emission would make the star colder than it
is. In particular, the core cannot contain any wide layer of
superfluid neutrons.

Moreover, even if the standard neutrino emission due to the modified
Urca (MU) process operated in the stellar core, the star would have
been insufficiently hot. This is shown by the solid cooling curve MU
in Fig. \ref{f:cool} and demonstrated as model MU in Table
\ref{tab:XMMU}. The star is supposed to have the standard heat
blanketing envelope made of iron (with a thin carbon atmosphere on
top). According to Fig. \ref{f:cool} and Table \ref{tab:XMMU}, the
MU model cannot explain the observations of \XMMU. The model implies
that the star is non-superfluid; in addition to the MU processes,
weaker processes of neutrino emission in neutron-neutron (nn),
neutron-proton (np) and proton-proton (pp) collisions (called
neutrino-pair bremsstrahlung in nucleon-nucleon
collisions) also operate
in the core but do not affect the cooling.

The neutrino emission of the star can be further reduced by strong
proton superfluidity in the stellar core. It greatly suppresses the
neutrino reactions involving protons which are the MU-process and
the weaker neutrino processes of pp and np collisions.
However, the neutrino
emission of non-superfluid neutrons  in nn collisions survives; it
is 30--100 times weaker than the MU process. The cooling scenario SF
is for the star with strong proton superfluidity and iron heat
blanket. The star becomes noticeably hotter than the MU star, but
still insufficiently hot to explain the data.

(ii) The second powerful regulator of the \XMMU\ surface temperature
is the amount of carbon (generally, of accreted matter containing
light elements -- \citealt{Potekhin_etal97}) in the heat blanketing
envelope. The thermal conductivity of the carbon envelope is higher
than that of the iron one; at the same internal temperature the
surface temperature becomes higher. Model MUac is for a
non-superfluid star with nearly maximum possible mass of carbon
$\DMC$ in the envelope ($\DMC \sim 10^{-8}M$). The surface
temperature of a non-superfluid star becomes noticeably higher than
that for the MU model, but it is still not high enough. Finally,
model SFac is for the superfluid star with the maximum mass of
carbon in the envelope. The double effect -- of proton superfluidity
and carbon envelope -- makes the surface temperature exceptionally
high (Fig. \ref{f:cool}, Table \ref{tab:XMMU}), in agreement with
observations.

Notice that the internal temperature of \XMMU\ is almost independent
of the composition of the heat blanket; in particular,
the $\Tg(t)$ curves MU and MUac in
Fig.\ \ref{f:cool} almost coincide at $t\sim$ 100 yr -- 100 kyr,
as well as the curves SF and SFac. This
is because \XMMU\ cools via neutrinos from inside. Accordingly,
the presence of carbon in the heat blanket just increases $\Ts$  making
the surface hotter without any back reaction on the neutron star interiors.

It is worth to mention that the cooling scenarios SF and SFac
can also be realized if direct Urca process is formally allowed in
\XMMU\ for a given EOS but is exponentially suppressed by very
strong proton superfluidity (similar situations are described, e.g.,
by \citealt{YP04}). 

We analyse the \XMMU\ cooling below. In two appendices we
discuss the parameters of heat blanketing envelopes and the neutrino
emission of the neutron star crust.

\section{Neutrino cooling function of neutron stars}
\label{s:coolfun}

Let us outline calculation of cooling curves,  $T_s^{\infty}(t)$,
for analyzing the \XMMU\ cooling. The surface temperature $T_s^{\infty}$
is directly related to the internal temperature
\citep{Yakovlev_etal11}.

The decrease of the internal temperature of a neutron star at the
neutrino cooling stage after reaching the internal thermal
relaxation is described by the well known equation (e.g.,
\citealt{Yakovlev_etal11})
\begin{equation}
  \frac{{\rm d}\widetilde{T}}{{\rm d}t}=-\ell(\widetilde{T}),\quad
  \ell(\widetilde{T})=\frac{L_\nu(\widetilde{T})}{C(\widetilde{T})}.
\label{e:cool}
\end{equation}
Here, $\widetilde{T}=T \sqrt{g_{00}}$ is the redshifted temperature
of the isothermal region within the star (it is this temperature
which is constant in isothermal regions in the frame of General
Relativity); $T$ is the local (non-redshifted) temperature, $g_{00}$
is the metric component in a given point of the star; $L_\nu$ and
$C$ are, respectively, the redshifted integral neutrino luminosity and
heat capacity of the star determined mainly by the super-dense core, and
$\ell(\widetilde{T})$ is the neutrino cooling function (e.g.,
\citealt{Yakovlev_etal11}). This function regulates the internal
cooling at the neutrino cooling stage.

\begin{figure}
\centering
\includegraphics[width=0.48\textwidth]{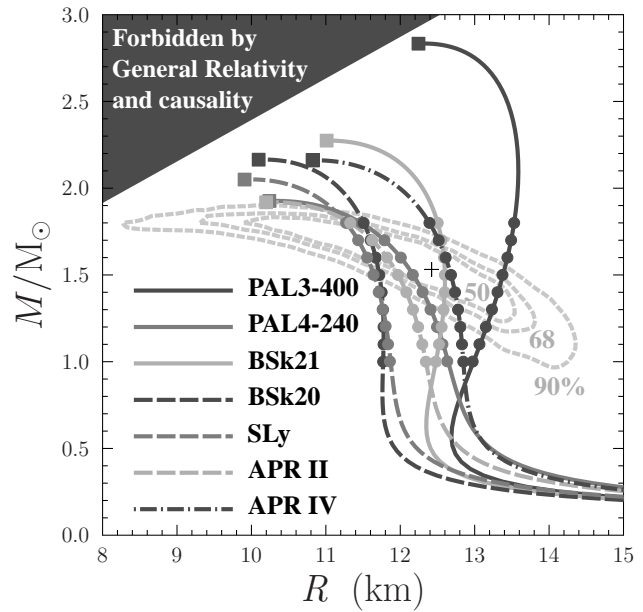} 
\caption{$M-R$ relations for neutron star models with the selected EOSs.
Filled squares refer to maximum mass models. Filled dots indicate
neutron star models used for calculating $q_{\rm MU}$ and $q_{\rm
SF}$ and for deriving the fit expressions (\ref{Eq_f-PS}) and
(\ref{Eq_f-DO}). Dashed contours are confidence regions for \XMMU\
obtained in Paper I at 50, 68 and 90 per cent  significance levels
by fitting the observed spectra with the carbon atmosphere models. The
cross is the best fit. The shaded upper left corner is prohibited by
General Relativity and causality. See text for details. }
\label{f:eoses}
\end{figure}

We have calculated $\ell(\Tg)$ for two cases. In the first case, the
star is assumed to be non-superfluid and cools mainly via the MU
process; such stars can be called standard neutrino candles. In the
second case, the star possesses very strong proton superfluifity
with the critical temperature for this superfluidity $\Tc \gtrsim 5
\times 10^9$ K everywhere in the core. Such superfluidity
completely suppresses the MU, np and pp processes of neutrino
cooling so that the star cools via the neutrino emission in nn
collisions. In addition, strong proton superfluidity fully
suppresses the proton heat capacity. In both cases $L_\nu \propto
\widetilde{T}^{~8}$ and $C \propto \widetilde{T}$. Then
\begin{equation}
\label{Eq_cooling1}
\ell(\widetilde{T}) = q\,\Tg_9^{~7}.
\end{equation}
The factor $q$ (to be expressed in K~s$^{-1}$) depends on $M$, $R$
and EOS in the stellar core (see below); $\widetilde{T}_{9} =
\widetilde{T}/(10^9\,\text{K})$.

The cooling equation (\ref{e:cool}) with the neutrino cooling function
(\ref{Eq_cooling1}) after reaching the state of internal thermal
relaxation of the star gives the well known analytic solution (e.g.
\citealt{Yakovlev_etal11})
\begin{equation}
\label{Eq_Tg-t}
\widetilde{T}_{9} = (6qt/T_*)^{-1/6},
\end{equation}
where $t$ has to be expressed in seconds; we insert the
normalization temperature $T_*=10^9$~K to ensure proper dimensions
of $q$ and $t$.

The factor $q=q(M,R)$ depends on proton superfluidity in a neutron
star core. We will see that the factors for non-superfluid stars
($q=q_{\rm MU}$) and for the stars with very strong proton superfluidity
($q=q_{\rm SF}$) are related as  $q_{\rm SF} \sim (0.01-0.02)\,
q_{\rm MU}$.  In reality, proton superfluidity only partly
suppresses the neutrino emission in reactions involving protons.

Following \citet{Yakovlev_etal11} we write
\begin{equation}
   q=q_{\rm MU}f_\ell,
    \quad f_\ell \equiv \ell(\Tg)/\ell_{\rm MU}(\Tg),
\label{e:fl}
\end{equation}
where $f_\ell$ is the neutrino cooling function of the star
expressed in terms of $\ell_{\rm MU}(\Tg)$
for a standard neutrino candle (Paper I); it is nearly temperature
independent and small (for \XMMU\ in our model). Evidently, we have
$q \geq q_{\rm SF}$. It is instructive to rewrite (\ref{e:fl}) in
the form
\begin{equation}
\label{Eq_reductDef}
q = \fl \, q_{\rm MU} + q_{\rm SF}.
\end{equation}
In this case, $\fl=f_\ell-\fln$, with $\fln \equiv q_{\rm SF}/q_{\rm
MU}$ being the minimum neutrino cooling function of \XMMU\ expressed
in standard neutrino candles. Thus defined, $f_{\ell \rm p} $
describes an incomplete suppression of the standard neutrino candle
emission by proton superfluidity. Such a suppression is determined
by the density profile of the critical temperature for proton
superfluidity $T_{\rm cp}(\rho)$ in the stellar core. Both
quantities, $T_{\rm cp}(\rho)$ and $f_{\ell \rm p}$, are a priori
unknown.

Equations (\ref{e:fl}) and (\ref{Eq_reductDef}) give two equivalent
methods to describe the effect of strong proton superfluidity on
neutron star cooling:
\begin{enumerate}

\item
One can use (\ref{e:fl}) and describe the effect of proton
superfluidity by the factor $f_\ell$ which is restricted ($f_\ell
\geq \fln$) by the neutrino cooling function due to non-superfluid
neutrons. This approach was used in Paper I although $\fln$ was not
accurately determined there. In principle, $f_\ell$ might also be
restricted by the neutrino emission from the neutron star crust but
the latter restriction is insignificant (Appendix \ref{appendix2}).

\item

Alternatively, one can employ (\ref{Eq_reductDef}) and characterize the
effect of proton superfluidity by the factor $\fl$ defined in such a
way that $\fl \to 0$ in the limit of very strong superfluidity. We
adopt this approach here.

\end{enumerate}
Notice that this refinement of the cooling theory is required only
for very slowly cooling neutron stars ($f_\ell \ll 1$). For other stars,
$f_\ell$ is much higher than $\fln$ so that $\fln$ can be
disregarded.

\begin{figure*}
\includegraphics[width=0.33\textwidth]{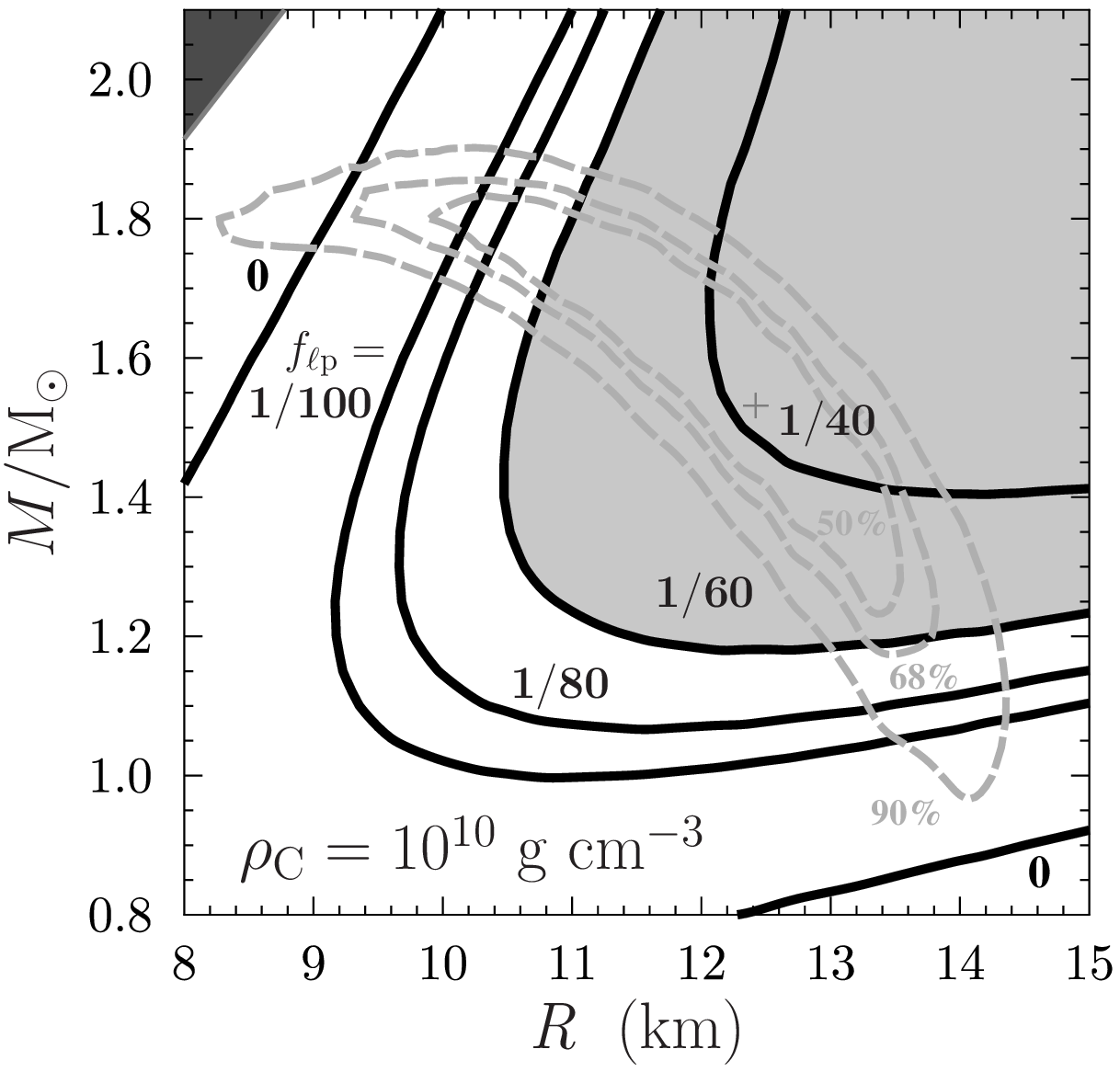}
\includegraphics[width=0.33\textwidth]{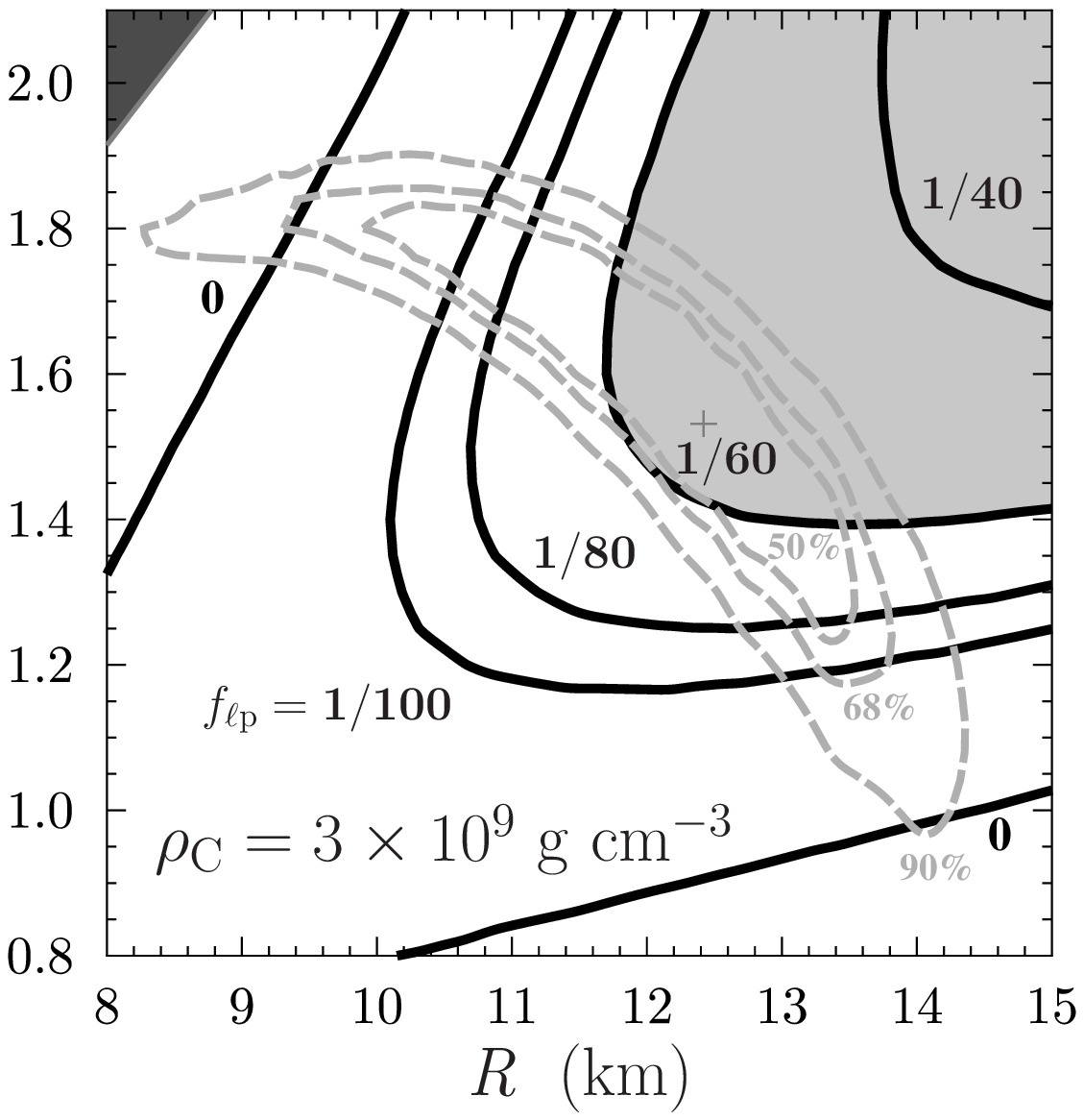}
\includegraphics[width=0.33\textwidth]{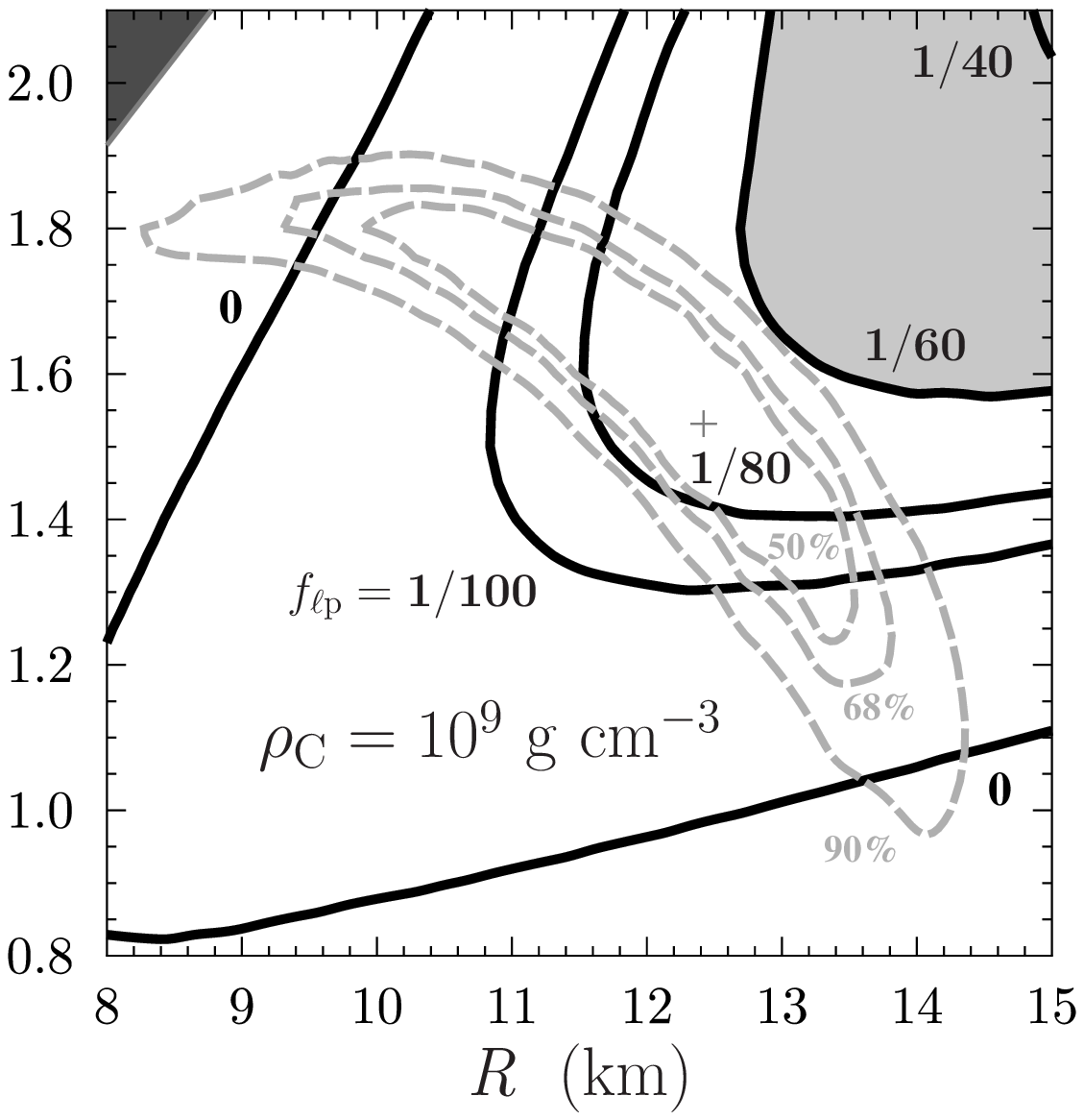}   
\caption{Solutions of the \XMMU\ cooling problem in the $M-R$ plane. The
lines correspond to fixed values of the suppression factor $\fl$=0,
1/100, 1/80, 1/60, and 1/40 of the neutrino cooling rate by proton
superfuidity for the models of heat blanketing envelopes containing
carbon up to the density $\rho_{\rm C}=10^{10}$ (left), $3 \times
10^9$ (middle) and $10^9$~g~cm$^{-3}$ (right). Light shading shows
the regions where $\fl \geq 1/60$.  The light dashed contours are
the same as in Fig.\ \ref{f:eoses}.
See text for details.}
\label{fig:contours}
\end{figure*}

In any case, (i) or (ii), we need $q(M,R)$ in the two limits, for
fully non-superfluid stars ($q=q_{\rm MU}$) and for the stars with
very strong proton superfluidity in the core ($q=q_{\rm SF}$). We
have calculated $q_{\rm MU}$ and $q_{\rm SF}$ for seven EOSs of
superdense matter in neutron star cores. The SLy, PAL4-240 and
PAL3-400 EOSs are described by \citet{Yakovlev_etal11}, APR II is
described by \citet{GKYG05}; BSk20 and BSk21 by
\citet{Potekhin_etal13}, and APR IV EOS by \citet{KKPY14} (who
called it the HHJ EOS). The $M(R)$ relations for neutron star models
with these EOSs are plotted in Fig.\ \ref{f:eoses}. The majority of
these EOSs cover the range of $M$ and $R$ values which is usually
treated as realistic \citep{HPY07}. However, some of the selected
EOSs (e.g. PAL3-400) are purely phenomenological and are thought to
be less realistic. They are included to enlarge the range of neutron
star radii $R$ involved in our analysis. Squares in Fig.\
\ref{f:eoses} refer to most massive stable neutron star models. One
can see that the chosen EOSs are reasonably consistent with recent
observations of two massive ($M \approx 2\,{\rm M_\odot}$) neutron
stars \citep{Demorest_etal10,Antoniadis_etal13}. The calculations of
$q=q_{\rm MU}$ and $q=q_{\rm SF}$ have been performed for a range of
masses $M$=1.0, 1.1, 1.2,\ldots 1.8 ${\rm M_\odot}$; 63 calculated
models are shown by filled dots. The majority of the selected EOSs
open direct Urca process in sufficiently massive stars. While
computing $q(M,R)$ we have artificially switched off fast neutrino
cooling due to direct Urca process.  This allows us to include
the case of switched on direct Urca process when it is exponentially
suppressed by strong proton superfluidity (Section \ref{sec:intro}).
In this case $q_{\rm MU}$ determines a formal standard-candle
neutrino emission level, our convenient unit for measuring the real
level in hot stars like \XMMU. Notice that the values $q(M,R)$ are
calculated using the same effective masses of nucleons and matrix
elements of neutrino reactions in neutron star cores which were used
by \citet{Yakovlev_etal11}.

We have approximated numerical values of $q_{\rm MU}$ by the
expression
\begin{equation}
\label{Eq_f-PS}
  q_{\rm MU}(M,R)=4.59\,\mathrm{K~s}^{-1}\,
    \frac{\gamma^{10}}{1+0.3 \gamma}\,\exp \left(0.16\,\frac{\beta}{x} \right),
\end{equation}
where
\begin{equation}
\label{Eq_Gamma-rg-rho}
 \gamma =
\frac{1}{\sqrt{1-x}},\quad x=\frac{r_{\rm g}}{R},
 \quad \beta =
\frac{3M}{4\pi R^3 \rho_0},
\end{equation}
$r_{\rm g}=2GM/c^2=2.95\, {M}/{\rm M_{\odot}}$~km is the
Schwarzschild radius and $\rho_0 = 2.8\times 10^{14}$ g~cm$^{-3}$ is
the density of saturated nuclear matter. The root mean square (rms)
relative fit error is 0.12, and the maximum relative error is 0.22
for the $M=1.8\,{\rm M_\odot}$ neutron star with the SLy EOS.

For the case of extremely strong proton superfluidity we have
derived a similar approximation,
\begin{equation}
\label{Eq_f-DO}
q_{\rm SF}=  0.174\; \text{K~s}^{-1} \frac{\gamma^5 \,
\exp\,[0.624\,( \gamma - 1 ) ( 1 + 1.03\,\beta )]}{1 + 0.0146
\,\beta}.
\end{equation}
Here the rms  relative fit error is 0.016 and
the maximum  relative error 0.035 takes
place for the $1.8\,{\rm M_\odot}$ star with the BSk21 EOS. Notice
that $q_{\rm MU}$ was approximated earlier by
\citet{Yakovlev_etal11}. Their fit is somewhat less accurate but
covers larger range of masses, $1 \leq  M/{\rm M_\odot}\leq 2.4$.
In our notations, their rms relative fit error of $q_{\rm MU}$
is  0.21, and the maximum error is 0.42. We have checked that
for $10 \leq R \leq 14$ km and $1
\leq  M/{\rm M_\odot}\leq 1.8$ the rms relative deviation of $q_{\rm
MU}$ values provided by the previous
and new fits is about 0.24, and the maximum
deviation $\sim 0.56$ is reached at $M=1.0\,{\rm M_\odot}$ and $R=14$ km.
As follows from a discussion below, such an agreement is quite
satisfactory. We have constructed a new fit to make the fitting of
$q_{\rm MU}$ and $q_{\rm SF}$ more uniform and to reduce the fit
error. We stress that both approximations, (\ref{Eq_f-PS}) and
(\ref{Eq_f-DO}), are valid for all selected EOSs. In this sense they
are universal. The fit errors reflect deviations from universality.

The applicability of our fits (\ref{Eq_f-PS}) and (\ref{Eq_f-DO}) is
greatly affected by a strong temperature dependence of the neutrino
cooling function (\ref{Eq_cooling1}). If one uses (\ref{Eq_Tg-t}) to
calculate the evolution of the internal stellar temperature
$\widetilde{T}(t)$, the relative errors  will be pretty small, about
6 times smaller than the indicated errors of the $q$-fits. The error of
calculating the surface temperature $\Ts$ would be additionally
twice smaller  because $\Ts$ approximately behaves as $\Ts \propto
\widetilde{T}^{1/2}$ \citep{GPE83}. This is a well known property of
the cooling theory: one can accurately calculate the neutron star
temperature with not very accurate cooling functions. The
temperature evolution appears to be really universal, almost
independent of the EOS. The prize for that is also well known: even
slight variations of the temperature correspond to sufficiently
large variations of the neutrino cooling rates (e.g.,
\citealt{Weisskopf_etal11}). In other words, it is difficult to
accurately determine the cooling rate from the data on the temperature.

In Fig.~\ref{f:cool} the dependence (\ref{Eq_Tg-t}) corresponds to
slightly bent straight segments of the cooling curves. The neutrino
cooling rate (\ref{Eq_f-PS}) is responsible for the MU and MUac
curves, whereas the rate (\ref{Eq_f-DO}) for the SF and SFac curves.
Equations (\ref{Eq_f-PS}) and (\ref{Eq_f-DO}) approximately describe
the evolution of the internal temperature $\widetilde{T}$ in
accordance with (\ref{Eq_Tg-t}) (the lower and upper dotted
$\Tg$-lines, respectively). In order to find the observed surface
temperature $\Ts$ one should use the relation between the surface
and internal temperatures of the star
\citep{Potekhin_etal97,Yakovlev_etal11}.

\section{Cooling of XMMU 1732}
\label{s:cooling}

In Paper I the observations of \XMMU\ were interpreted using the carbon
atmosphere models of neutron stars. For a wide range of possible
masses  $M$ and radii $R$ of neutron stars the observed spectra of
\XMMU\ were fitted with these theoretical models and the effective
surface temperature  $\Ts$ was determined. For  $d=$3.2 kpc, the
confidence contours of $M$ and $R$ at 50, 68 and 90 per cent
significance levels are plotted in Fig.\ \ref{f:eoses}. The cross
shows the best fit which corresponds to $M=1.53\,{\rm M}_\odot$,
$R=12.4$~km and $\Ts=1.78$~MK. The confidence contours restrict
allowable masses and radii of \XMMU\ for the assumed distance.

Let us show that the cooling theory further restricts allowable
values of $M$ and $R$. To this aim, for every pair of $M$ and $R$ we
have constructed theoretical cooling curves using equations
(\ref{Eq_Tg-t}) and (\ref{Eq_reductDef}) as well as the relation
between the internal and surface temperatures calculated by A.
Potekhin \citep{Yakovlev_etal11}. The heat blanketing envelope is
assumed to extend to the density $\rho_{\rm b}=10^{10}$~g~cm$^{-3}$,
and has the outside carbon layer (to a density $\rhoC$) and an
underlying layer of iron. In our case, the cooling is regulated by
four parameters $M$, $R$, $\fl$ and $\rhoC$. Here we use $\rhoC$,
which is more convenient than $\DMC$ used in Paper I. The relation
between $\DMC$ and $\rhoC$ is discussed in Appendix \ref{appendix1}.

Now we can employ equation (\ref{Eq_Tg-t}), combine it with the
relation between the surface and internal temperatures, and
calculate $\Ts$ for the assumed age of $t=27$ kyr. The presented
formalism makes these calculations very simple, fast and
straightforward (no need to run a sophisticated cooling code). For
any pair of $\fl$ and $\rhoC$, we can now immediately find families
of values of $M$ and $R$ which give the surface temperature $\Ts$
inferred in Paper I from the interpretation of the spectra. These
values of $M$ and $R$ give us possible solutions of the reciprocal
cooling problem for \XMMU. Such solutions are independent of the EOS
in the stellar core because they are obtained with the universal
approximations (\ref{Eq_f-PS}) and (\ref{Eq_f-DO}).

Fig.~\ref{fig:contours} shows several curves on the $M-R$ diagram
which visualize the solutions of the reciprocal cooling problem. The
left-hand panel is for the most massive, fully carbon heat
blanketing envelope with $\rhoC = \rho_{\rm b}=
10^{10}$~g~cm$^{-3}$. Each curve corresponds to a fixed value
$\fl$=0, 1/100, 1/80, 1/60 and 1/40. For weaker proton superfluidity
(higher $\fl$) these curves shift to higher $M$ and $R$. The middle
panel shows the cooling solutions for the same values of $\fl$, but at
the smaller amount of carbon in the heat blanket, $\rhoC = 3\times
10^{9}$ g~cm$^{-3}$. At any fixed $\fl$ the decrease of $\rhoC$ also
shifts the curves to higher $R$ and $M$.
The righ-hand panel is again for the same $\fl$ but for still
smaller amount of carbon, $\rhoC=10^9$~g~cm$^{-3}$. The curves of
constant $\fl$ continue shifting to higher $M$ and $R$. Note that
the curves in Fig.\ \ref{fig:contours} are very sensitive to
the approximations of $q_{\rm MU}$ and $q_{\rm SF}$. This is a direct
consequence of strong temperature dependence of the neutrino cooling
function $\ell(\Tg)$ as detailed above.

The formal solutions of the cooling problem
(Fig.~\ref{fig:contours}) have to be reconciled with our knowledge
of general properties of neutron star matter. The crucial parameter
in our model is $\fl$. It is determined by the profile of critical
temperature $\Tc(\rho)$ for the onset of proton superfluidity in the
neutron star core. As mentioned above, the limiting value $\fl=0$
corresponds to very strong proton superfluidity, with
$\Tc(\rho)\gtrsim 5 \times 10^9$~K everywhere in the core. Since
theoretical values $\Tc(\rho)$ are very model dependent \citep{LS01}
we would not like to rely on any specific model for proton
superfluidity. Nevertheless, it is clear on theoretical grounds that
the condition  $\Tc(\rho)\gtrsim 5 \times 10^9$~K in the entire core
is unrealistic, especially in central regions of massive stars where
the density of the matter is especially high. At very high
densities,  nuclear attraction of protons
will inevitably turn into repulsion which should destroy proton
superfluidity, increase the neutrino emission,
and noticeably cool \XMMU, in disagreement with observations.

To summarize this discussion, very small values $\fl \to 0$ seem
unrealistic, especially in massive stars. Realistic minimum values
of $\fl$ have to be determined from a careful analysis of
$\Tc(\rho)$ calculations for the different models of nucleon
interactions and various models of neutron stars which is beyond the
scope of this paper. By way of illustration, we assume that $\fl
\gtrsim 1/60$. As seen from Fig.\ \ref{fig:contours}, this
assumption would immediately impose serious constraints on mass and
radius of \XMMU. For the fully carbon heat blanketing envelope
(left-hand panel) we would have $R>10.5$ km and $M> 1.2\,{\rm
M_\odot}$. For slightly less amount of carbon ($\rhoC=3 \times 10^9$
g~cm$^{-3}$, middle panel) we have $R>12$ km and $M>1.37\,{\rm
M_\odot}$. For smaller amount of carbon ($\rhoC=10^9$ g~cm$^{-3}$,
right-hand panel) the allowable values of $M$ and $R$ become
uncomfortably high, barely compatible with the values derived from
the spectral fits (Fig.~\ref{f:eoses}). Therefore, it seems that the
heat blankets with $\rhoC \lesssim 10^9$ g~cm$^{-3}$ are not favored
by the cooling models of \XMMU, in agreement with the qualitative
consideration of Paper I.

\begin{figure}
\centering
\includegraphics[width=0.45\textwidth]{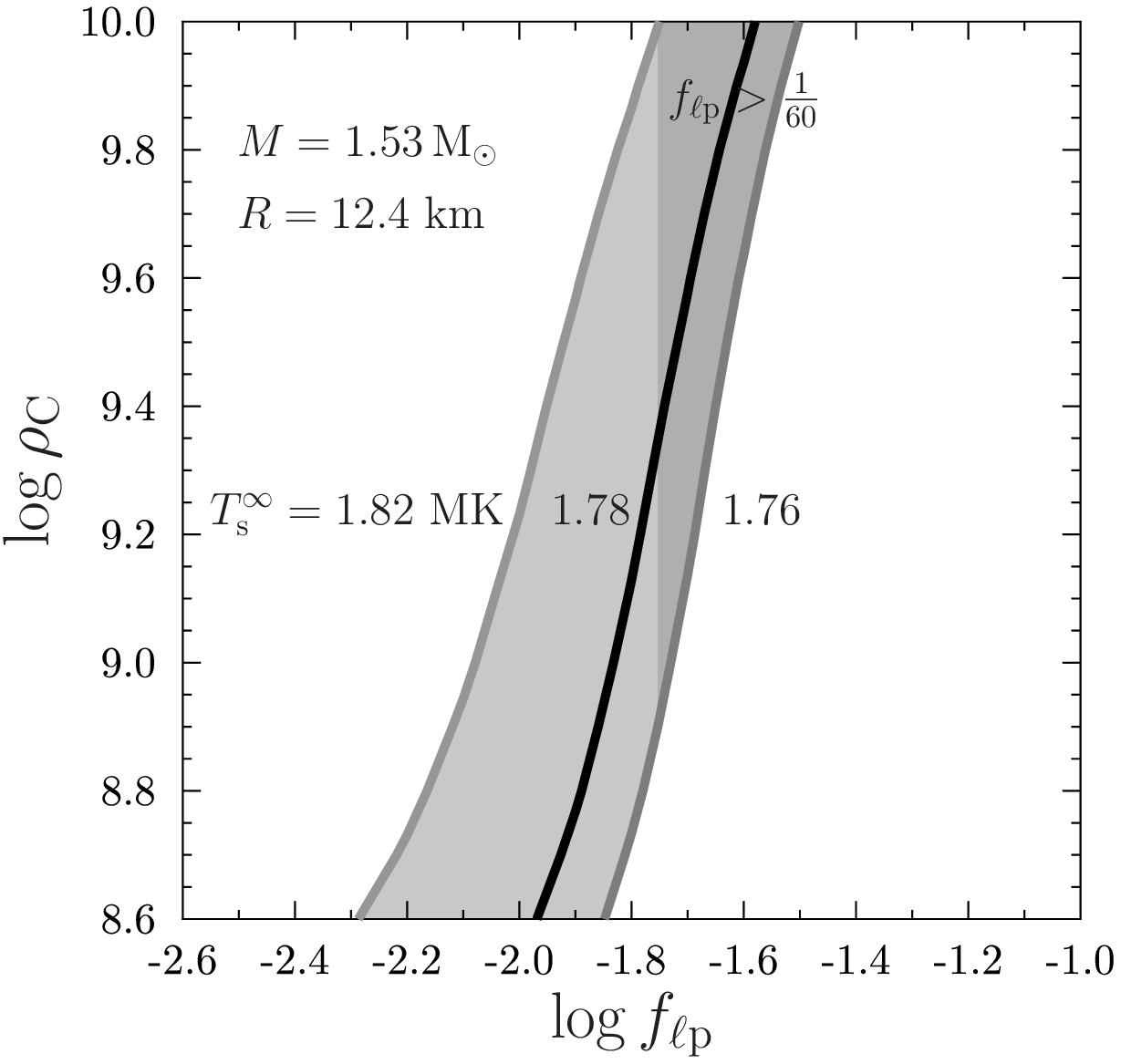} 
\caption{Solutions of the \XMMU\ cooling problem in the $\fl-\rhoC$
plane for the star of $M=1.53\,{\rm M_\odot}$ and $R=12.4$ km with
the measured surface temperature
$\Ts=1.78^{+0.04}_{-0.02}$ MK. The central thick black line
corresponds to the best-fit surface temperature $\Ts=1.78$ MK, while
the other two lines are for $\Ts=1.76$ and $\Ts=1.82$ MK. Weakly shaded
is the formal allowable range of $\fl$ and $\rhoC$. The densely
shaded is a more realistic allowable range restricted by $\fl\geq
1/60$. See text for details.}
\label{f:bestfit}
\end{figure}

Let us add that according to numerous calculations and theoretical
analysis of observations (e.g., \citealt{Klae2006,BY2015} and
references therein), the powerful direct Urca process of neutrino
cooling really opens in sufficiently massive neutron stars.
If it is not suppressed by strong superfluidity, it will
not allow these stars to be as hot as \XMMU. From this point of view
one can expect that \XMMU\ is not very massive (say, $M \lesssim
1.6\,M_\odot$). Accordingly, one can further reduce the $M$--$R$
range allowed by cooling models (shaded regions in Fig.
\ref{fig:contours}) by removing the region of rather high masses.
However, this additional restriction of the cooling models requires
further consideration. Generally, cooling theories predict that
massive stars cannot be very hot.  In principle, the direct
Urca process can be suppressed by strong superfluidity but
the existence of strong superfluidity in central parts of
massive stars is unlikely on theoretical grounds (see above). However,
\XMMU\ can be a moderate-mass star, with the direct Urca process
formally allowed but exponentially suppressed by strong superfluidity.

\begin{figure*}
\centering
\includegraphics[width=0.48\textwidth]{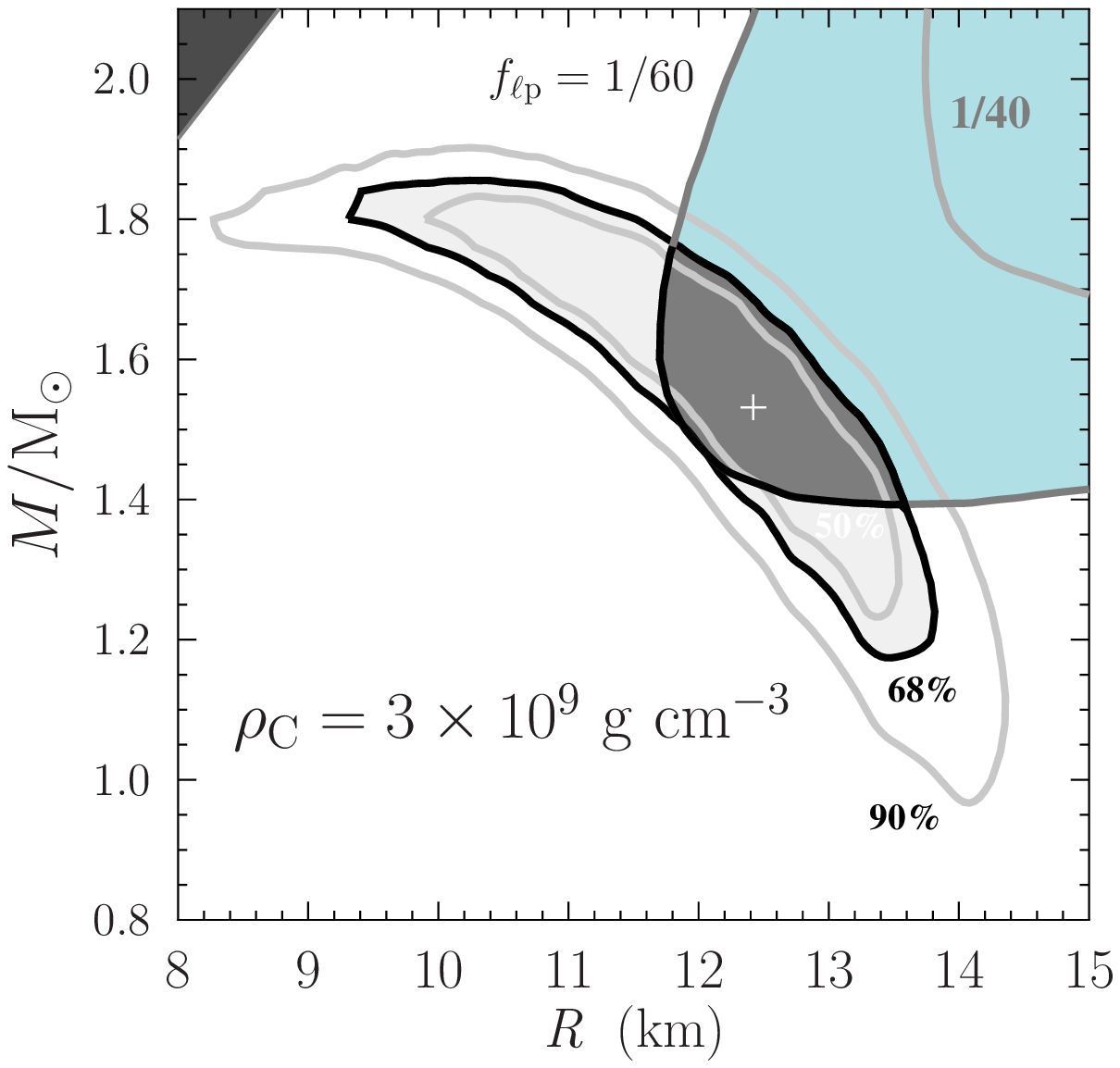}
\includegraphics[width=0.48\textwidth]{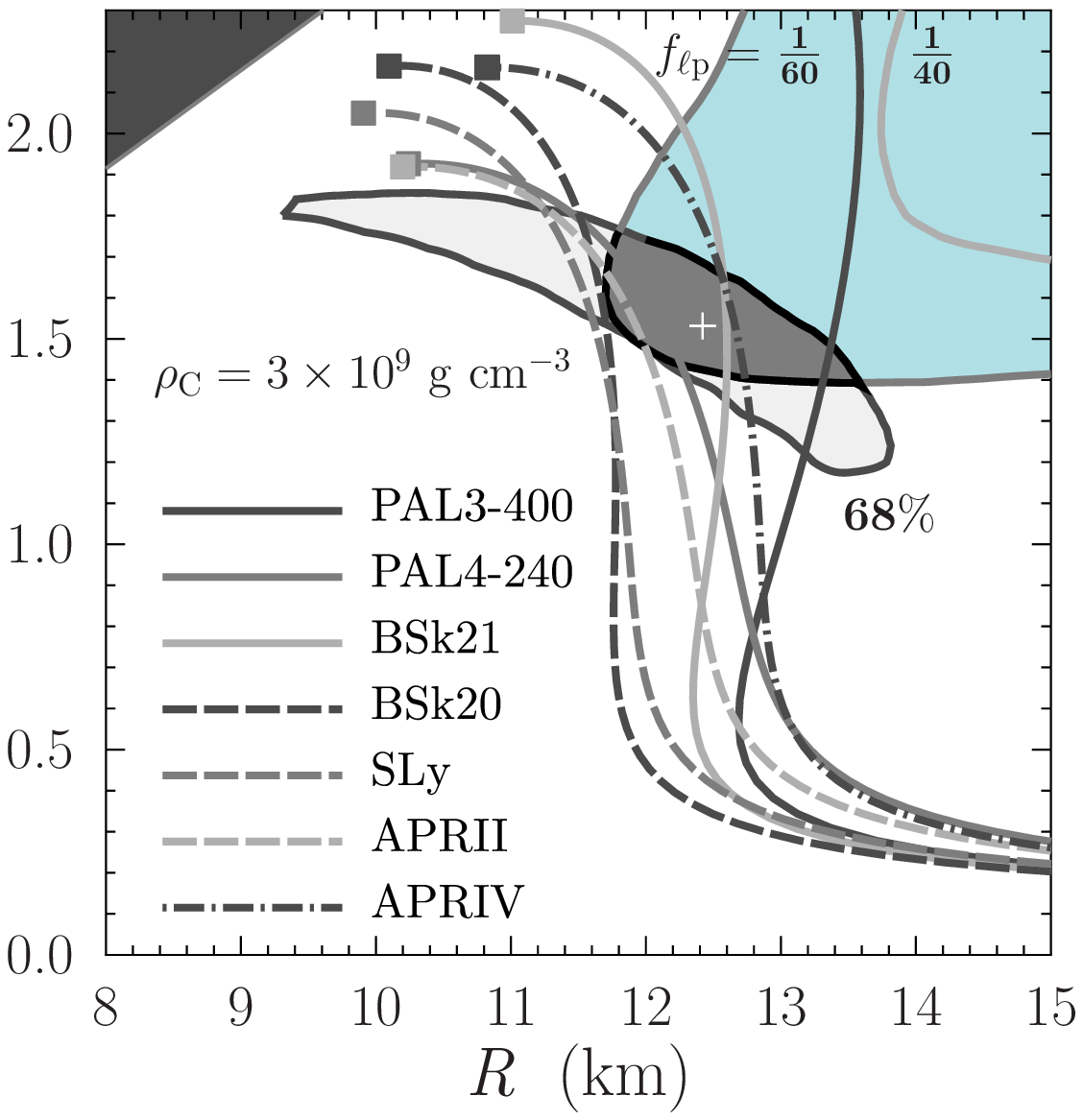}
\caption{Constraining $M$ and $R$ of \XMMU\ from fitting the
observed spectra (Fig.\ \ref{f:eoses}) and the cooling theory (Fig.\
\ref{fig:contours}). We employ the 68 per cent confidence region
given by the spectral fits. As far as the cooling theory is
concerned, we assume the carbon envelope with $\rhoC=3 \times
10^9$~g~cm$^{-3}$ and $\fl>1/60$. Densely shaded is the resulting
confidence $M-R$ region. Left-hand panel: same $M-R$ scales as in
Fig.\ \ref{fig:contours}, with the 50 and 90 per cent contours obtained
from the spectral fits also shown. Right-hand panel: larger $M$ range;
the mass-radius relations for neutron stars with the different EOSs (Fig.
\ref{f:eoses}) are added. See text for details.} \label{f:obstheor}
\end{figure*}

Let us recall that Paper I used the alternative description for
the superfluid suppression of the neutrino cooling rate (by the
parameter $f_\ell$ instead of the parameter $\fl$ used here; see
Section \ref{s:coolfun}). The red dashed curve in fig.\ 8 of Paper I
corresponds to $f_\ell=1/40$. We have checked that it is
qualitatively similar but not exactly coincident with the curve
$f_\ell=1/40$ calculated using our current formalism at $\rhoC=3
\times 10^9$~$\gcc$. The nature of the difference is twofold.
Firstly, the dashed curve in Paper I is calculated for a constant mass
fraction of the carbon envelope, $\DMC/M$, whereas here we take
constant $\rhoC$. Secondly, here we employ the refined fit
(\ref{Eq_f-PS}) for $q_{\rm MU}$, whereas Paper I used slightly less
accurate fit taken from \citet{Yakovlev_etal11}.

Finally, Fig.\ \ref{f:bestfit} illustrates the sensitivity of
extracting the neutrino cooling rate (the parameter $\fl$) from a
measured surface temperature $\Ts$ of the star. To this aim, we have
selected the best-fit neutron star model ($M=1.53\,{\rm M_\odot}$,
$R=12.4$ km; crosses in Figs. \ref{f:eoses} and \ref{fig:contours})
and assumed that the measurements give
$\Ts=1.78^{+0.04}_{-0.02}$~MK. The solid black central line gives
formal solutions of the cooling problem in the $\fl-\rhoC$ plane.
The line means the family of such solutions with different $\fl$ and
$\rho_C$ which give one and the same surface temperature $\Ts=1.78$ MK. All
of therm are formally allowed by our cooling models. Two thick gray
lines are similar solutions for the limiting temperatures $\Ts=1.76$
and 1.82 MK. Weakly shaded region between the gray lines is the
range of $\fl$ and $\rhoC$ which is formally allowed at
$\Ts=1.78^{+0.04}_{-0.02}$~MK. We see that a narrow interval of
$\Ts$ corresponds to rather wide intervals of $\fl$ and $\rhoC$. If
we assume, on physical grounds, that $\fl \geq 1/60$ (as in Fig.
\ref{fig:contours}), the allowable $\fl-\rhoC$ range would be much
more restricted (densely shaded region in Fig.\ \ref{f:bestfit}).

Now we can combine the mass-radius constraints given by spectral
fits (Fig.\ \ref{f:eoses}) and the cooling theory (Fig.\
\ref{fig:contours}). One example of such analysis is presented in
Fig.\ \ref{f:obstheor}. To be specific, we take the 68 per cent
banana-like confidence region of $M$ and $R$ from Fig.\
\ref{f:eoses}. For certainty, we assume the presence of carbon in
the heat blanketing envelope up to the density $\rhoC=3 \times 10^9$
$\gcc$ (the shaded region of $M$ and $R$ in the middle panel of
Fig.\ \ref{fig:contours}). The left-hand panel of Fig.\
\ref{f:obstheor} is plotted for the same ranges of $M$ and $R$  as
in Fig.\  \ref{fig:contours}. It shows also the 50 and 90 per cent
confidence contours from Fig. \ref{f:eoses}. The right-hand panel is
plotted for a larger range of masses and presents also the mass-radius
relations for the different EOSs from Fig. \ref{f:eoses}. As seen from
Fig.\ \ref{f:obstheor}, the cooling theory strongly reduces the
banana-like $M-R$ region. Let us mention that if we used only
realistic theoretical EOSs of neutron stars (e.g., \citealt{HPY07})
to reduce the banana-like region, we would do similar reduction. The
joint region allowed by the observed spectra and cooling models is
densely shaded in Fig.\ \ref{f:obstheor}. The cooling models
disfavor very low and very large radii (both ends on the banana),
just in accordance with theoretical expectations. Notice that  the
best fit $M$ and $R$ obtained from the spectral fits (the cross in
Figs.\ \ref{f:eoses}, \ref{fig:contours} and \ref{f:obstheor})
appear to be near the center of the densely shaded region.

Our analysis based on Fig.\ \ref{f:obstheor} is evidently not strict
because we have assumed specific values of $\rhoC=3 \times 10^9$
$\gcc$ and specific minimal $\fl=1/60$. We could slightly increase
$\rhoC$ (up to its maximum value $\rhoC=10^{10}$ $\gcc$) which would
widen the range of $M$ and $R$ provided by the cooling models at
$\fl \geq 1/60$  (left-hand panel on Fig.\ \ref{fig:contours}).
However, this limiting case of the heat blanketing envelope fully
composed of carbon seems not very realistic because at high densities
carbon will start burning in pycnonuclear reactions (especially in
carbon matter, containing admixture of other elements; see, e.g.,
\citealt{YGA2006}). On the other hand, if we slightly decrease
$\rhoC$ below $3 \times 10^9$ $\gcc$, the $M-R$ range allowed by the
cooling models with $\fl \geq 1/60$ will shift to higher $M$ and $R$
(outside the banana range, where, in addition, the existence of
exceptionally hot \XMMU\ is questionable due to possible opening of
the direct Urca process).  In principle, we could take the minimum
value of $\fl$ lower than 1/60. This would increase the $M-R$ range
allowed by the cooling models but very low $\fl$ seem unrealistic
(as explained above). All in all, although we have presented only
one example of chosen $\rhoC$ and $\fl$, we have actually not much
freedom to vary the parameters and be consistent with the
observations.

\section{Conclusions}
\label{s:conclude}

We have extended the cooling theory of neutron stars to study a very
slow cooling of exceptionally hot middle-aged stars. Such a cooling
can take place (e.g., Paper I) under the  effects of strong proton
superfluidity in stellar cores (to suppress internal neutrino
cooling) and large amount of sufficiently light elements in the
heat blanketing envelopes (to increase the heat transparency of
the envelopes and rise the surface temperature).

We have presented simple expressions (\ref{Eq_Tg-t}),
(\ref{Eq_reductDef}), (\ref{Eq_f-PS}) and (\ref{Eq_f-DO}) which
enable fast and accurate calculation of the internal temperature
$\Tg(t)$ of a very slowly cooling neutron star as a function of age
$t$ at the neutrino cooling stage. The expressions are obtained for
neutron stars with nucleon cores, where protons can be in a
superfluid state. The expressions are universal, valid for many EOSs
of nucleon matter. Equations (\ref{Eq_f-PS}) and (\ref{Eq_f-DO}) are
derived by fitting the results of numerical calculations for many EOSs.
Equation  (\ref{Eq_f-DO})  takes into account a very slow (but finite)
neutrino cooling in the presence of strong proton superfluidity
which has not been calculated earlier. The effect of proton
superfluidity is described by one parameter $\fl$ which is the
suppression factor of the neutrino cooling of the star with respect
to  a standard
neutrino candle. In our case the cooling is regulated by
neutron star mass $M$, radius $R$, factor $\fl$, and by
the amount of light elements in the heat blanketing envelope.

The advantage of our approach is that it is universal  and does not
depend on a specific model for proton superfluidity. All the
information on proton superfluidity is contained in $\fl$. It is
$\fl$ which can be inferred from observations; allowable models of
$\Tc$ to ensure this $\fl$ can be analyzed later.  These
models can describe the cases in which a non-superfluid star
would cool mainly via modified or even direct Urca process but
both Urca processes are greatly suppressed by proton superfluidity.
In this sense our
consideration extends model-independent analysis of cooling neutron
stars with standard cooling function $\ell(\widetilde{T}) \propto
\widetilde{T}^{~7}$, started by \citet{Yakovlev_etal11} and
\citet{Weisskopf_etal11}, and a more complicated model-independent
analysis of the cooling enhanced by the onset of triplet-state pairing
of neutrons and associated neutrino emission in the neutron star
core \citep{SY2015}.

The cooling model has been applied to interpret the observations of
thermal radiation of the \XMMU\ neutron star in the supernova
remnant HESS J1731--347. A preliminary interpretation was presented
in Paper I. Following Paper I we have assumed the carbon atmosphere
models, the distance $d=3.2$~kpc, and the neutron star age $t=27$~kyr. The
modified theory has noticeably improved the interpretation of
observations. We have obtained that the reasonable values of $\fl$
should be around 1/60 and the heat blanketing envelope should
contain a lot of carbon, up to the density $\rhoC \gtrsim 3 \times
10^9$~$\gcc$. The theory has allowed us to strongly restrict the
range of masses and radii of  \XMMU\ (see the densely shaded region in
Fig.~\ref{f:obstheor}) in comparison with the ranges obtained from
spectral fits.

Nevertheless we would like to warn the reader that that these
results can be considered as semi-quantitative. For instance,
strictly speaking,  the factor $\fl$ can vary with time (larger
layers of the core become superfluid). We have neglected this effect
assuming it is weak. Moreover, owing to the strong temperature
dependence of the neutrino cooling function $\ell(\Tg)$
(Section~\ref{s:coolfun}), the values
of $\fl$ which we infer from the observations are very sensitive to the
measured values of $\Ts$ and to a not very certain microphysics of the
neutron star matter. In particular, they are sensitive to the thermal
insulation of the heat blanketing envelopes (to the relation between the
surface and internal temperatures). Another example -- our equations
(\ref{Eq_f-PS}) and (\ref{Eq_f-DO}) are obtained using certain
expressions (e.g., \citealt{YKGH01}) for the neutrino emission in
the modified Urca process and nucleon-nucleon collisions (Table
\ref{tab:XMMU}). Although these expressions are widely used in
cooling simulations, they are model dependent. Were the theory of
neutrino processes improved (first of all, with regard to matrix
elements of the processes), equations (\ref{Eq_f-PS}) and
(\ref{Eq_f-DO}) should have been updated which may change the
results.

In addition, our consideration is based on the age of \XMMU\ equal
to 27~kyr. However, one cannot exclude that the age is different
which would affect the results. If the age were larger, say, 40 kyr,
our cooling model would still be able to explain the data but
assuming the strongest proton superfluidity and fully carbon blanketing
envelope. Were the age lower (e.g., as low as 10 kyr) the situation
would be more relaxed, than at $t=27$ kyr, but we would still need
both, strong superfluidity and massive carbon envelope. To become an
`ordinary' cooling neutron star (instead of extraordinary hot one)
its age should be $t\lesssim 3$ kyr. The distance to \XMMU\ is also
not very certain. Were $d=4.5$ kpc instead of 3.2 kpc, the values of
$M$ and $R$ inferred from the spectral fits would be noticeable higher,
not very realistic for neutron stars, and the inferred surface
temperature would also be slightly higher (Paper I). Although there
is no rigorous proof, it is widely believed that such a very massive
star should cool rapidly, in disagreement with the inferred $\Ts$.

\section*{acknowledgements}
The work of AK and DY was partly supported by Russian Foundation for
Basic Research (grants Nos. 14-02-00868-a and 13-02-12017-ofi-M) and
the work of VS by DFG grant WE 1312/48-1.


\appendix

\section{Parameters of Carbon Envelope}
\label{appendix1}

A carbon ($^{12}$C) layer in the heat blanketing envelope of a
neutron star can be characterized by the density $\rhoC$ at the
bottom of this layer. Alternatively, the layer can be specified by
the ratio $\Delta M_{\rm C}/M$ of the gravitational masses of the layer,
$\Delta M_{\rm C}$, and of the entire star, $M$. These quantities
are known to be related as (e.g., \citealt{Potekhin_etal97})
\begin{eqnarray}
  \frac{\Delta M_{\rm C}}{M}
  &=&\frac{P(\rhoC)}{P_0 g_{14}^2}
  = \frac{1.510 \times 10^{-11}}{g_{14}^2}
\nonumber \\
  &\times& \left[\xr \sqrt{1+\xr^2}
  \left(\frac{2}{3}\xr^2-1   \right)
  + \ln \left(\xr+\sqrt{1+\xr^2}  \right) \right],
\label{e:deltam}
\end{eqnarray}
where $g_{14}$ is the surface gravity in units of $10^{14}$
cm~s$^{-2}$, $P_0=1.193 \times 10^{34}$ dyn~cm$^{-2}$,
$\xr=1.009\,(\rho_6/\mu_{\rm e})^{1/3}$ is the relativistic
parameter of degenerate electrons, $\mu_{\rm e}=A/Z=12/6=2$ is the
number of nucleons per one electron in carbon matter,
$\rho_6=\rhoC/(10^6\,\gcc)$, and $P(\rhoC)$ is the pressure at the
bottom of the carbon layer which is approximated by the pressure of
free degenerate relativistic electrons; e.g., \citet{HPY07}. For a
star with $M=1.4\,{\rm M_\odot}$ and $R=12$ km ($g_{14}=1.59$,
$x=r_{\rm g}/R=0.345$) at $\rhoC=10^6$, $10^8$ and $10^{10}$ $\gcc$
one has $\DMC/M=8.680\times 10^{-13}$, $7.105 \times 10^{-10}$ and
$3.500 \times 10^{-7}$, respectively.

In addition, the layer can be characterized by the
column baryon mass density
of carbon, $\Sigma$, which is related to $\Delta M_{\rm C}$
as
\begin{equation}
   \Sigma = \frac{\Delta M_{\rm C}}{4 \pi R^2 \sqrt{1-x}}.
\label{e:Sigma}
\end{equation}
The dependence of $\Sigma$ and $\Delta M_{\rm C}/M$ on $\rhoC$ is, of course,
similar.

\section{Analytic approximation of neutrino luminosity of crust}
\label{appendix2}

Let us obtain an analytic approximation for the neutrino luminosity of
a neutron star crust in the reference frame of distant observer,
\begin{equation}
\label{e:Lcr}
  L_{\rm cr} = \int_{\rm crust}
  Q \,\exp(2\Phi)\,{\rm d}V,
\end{equation}
where $Q$ is a local neutrino emissivity, $\exp(2\Phi) = g_{00}(r)$
is a time-like component of the metric tensor, $r$ is a
circumferential radius, and d$V$ a proper volume element. The
integration is carried out over the crust volume.

Assume that the neutron star is thermally relaxed. Then the crust is
nearly isothermal; $\Tg$ is constant over the entire crust excluding a thin
outer heat blanketing envelope whose contribution to $L_{\rm cr}$ is
negligible. In this case the local temperature is given by $T(r) =
\Tg\,\exp(-\Phi)$. Because the crust is thin (with the thickness of
$\sim 0.1 R$) and contains $\sim 0.01$ of the neutron star mass, its
structure can be calculated in the relativistic Cowling
approximation; it is nearly independent of the EOS in the stellar
core. The metric function in the crust can be approximated as
$g_{00}=1-r_{\rm g}/r$, and the pressure gradient as
\begin{equation}
\label{e:TOV}
 \frac{{\rm d}P}{{\rm d}r}  =  -\frac{G\rho M}{r^2} \frac{1}{1-r_{\rm
 g}/r}.
\end{equation}

We replace the integration over $r$ in (\ref{e:Lcr})
by the integration over a
new variable $s$ defined as
\begin{equation}
\label{e:s}
    s(P)=\int_0^P \frac{{\rm d}P}{\rho c^2},
\end{equation}
where $P=P(r)$. Then one can obtain
\begin{equation}
\label{e:Lcr-s}
   L_{\rm cr} = 8\pi r_{\rm g}^3 \gamma^5 \int_{s_{\rm b}}^{s_{\rm cc}}
   \frac{Q \, \exp[-3(s - s_{\rm b})]}
   {\left\{ \gamma^2 - \exp[-2(s - s_{\rm b})] \right\}^4} \, {\rm
   d}s;
\end{equation}
$s_{\rm b}$ corresponds to the bottom of the heat blanketing envelope
($\rho=\rho_{\rm b}=10^{10}~\gcc $), and $s_{\rm cc}$ to the
crust-core interface ($\rho_{\rm cc}\approx 1.5 \times
10^{14}~\gcc$);
$\gamma$ is given by equation~(\ref{Eq_Gamma-rg-rho}).

Let us consider the most important case in which the main
contribution to $L_{\rm cr}$ comes from the neutrino-pair bremsstrahlung
due to collisions of strongly degenerate relativistic electrons with
atomic nuclei. The neutrino emissivity of this process has been
thoroughly investigated. Schematically, it can be written as $Q
\propto T^6 \Lambda(\rho, T)$, where $\Lambda$ is a Coulomb
logarithm (a weakly varying function of $T$ and $\rho$), while the
normalization factor depends only on $\rho$. In addition, the
Coulomb logarithm behaves smoothly at sufficiently high densities which
give the main contribution to $L_{\rm cr}$ (see, e.g.,
\citealt{Kam1999,Ofe2014}). With this in mind one can show that at
temperatures $\Tg \sim 3\times 10^7 - 10^9$~K of practical interest the neutrino
luminosity $L_{\rm cr}$ can be accurately approximated as
\begin{equation}
\label{e:Lcr1}
L_{\rm cr} = 9.053 \times 10^{34}~{\rm erg~s^{-1}}~\left(
\frac{M}{\rm M_\odot} \right)^3  \frac{\Tg_9^6 \gamma^{11}}{\left(
\gamma^2-1 \right)^4}\,\frac{\phi( y )}{\psi(z)},
\end{equation}
where $y=\Tg_9 \gamma/1.0643$, $z=\gamma^2-1$,
\begin{eqnarray}
 \phi(y) = ay\,\frac{(y-1)\exp(-p y)+1}{p+1}, & & a = 1.7, \; p =
 13.0;
\label{phiDef}\\
\psi(z) = \left( \frac{b}{z}-1 \right)\,\exp(-qz) + 1, & & b = 0.1, \;
q = 15.0.
\label{psiDef}
\end{eqnarray}
The approximation has been obtained using the smooth composition
model of the ground-state crust \citep{HPY07}.

We have compared the approximated $L_{\rm cr}$, equation
(\ref{e:Lcr1}), with that calculated numerically from equation
(\ref{e:Lcr-s}). The comparison has been made for the models of the neutron
star crust with the FPS or SLy EOS \citep{HPY07} but with the same
description of the nuclear composition as in (\ref{e:Lcr1}). We have
considered the wide ranges of masses $1.0\, {\rm M_\odot} \leq M \leq
2.5\, {\rm M_\odot}$, radii $10 \leq R \leq 16$ km and internal
temperatures $3 \times 10^7 \leq \Tg \leq 10^9$~K. The relative
deviations of approximated and numerical values of $\log L_{\rm cr}$
[erg~s$^{-1}$] do not exceed 2--3 per cent. In addition, we have
calculated $L_{\rm cr}$ for the 1.4$\,{\rm M_\odot}$ neutron star
model with the BSk21 EOS in the core and compared with $L_{\rm cr}$
given by equation (\ref{e:Lcr1}). The relative deviations turn out
to be as small as above. Furthermore, we have checked that the
neutrino emissivity due to the electron bremsstrahlung on atomic nuclei
in the crust made of BSk20 or BSk21 EOS taken with its proper
nuclear composition \citep{Potekhin_etal13} is almost the same as
for the smooth-composition model used in equation~(\ref{e:Lcr1}).

The above analysis demonstrates that our approximation is almost
universal. It is nearly independent of the EOS of superdense matter
in the neutron star core. Moreover, according to our calculations, the
main contribution to $L_{\rm cr}$ comes from the density range  $
10^{13}~\gcc~\lesssim \rho \leq \rho_{\rm cc}$. In this range the
models of accreted crust are almost indistinguishable
\citep{HPY07,Ofe2014} from the models of ground-state crust
(cold-catalyzed matter). Therefore, $L_{\rm cr}$ should be weakly
dependent of the EOS in the crust.

As \XMMU\ is thought to undergo extremely slow neutrino cooling, one
might think that the neutrino emission from its crust can be important
along with the neutrino emission due to nn-collisions in the core.
However, simple estimates based on equation (\ref{e:Lcr1}) show that
for the range of $M$ and $R$ of our interest (Section
\ref{s:cooling}) $L_{\rm cr}$ is typically by more than 10 times
lower than the neutrino luminosity due to nn-collisions in the core.
Therefore, it can be disregarded for the \XMMU\ cooling problem.

Nevertheless, one might speculate on the existence of very low-mass
neutron stars, with masses of $\sim (0.1-0.6)\, {\rm M_\odot}$.
These stars are hypothetical, difficult to produce on evolutionary
grounds. If, however, they exist, they would have large radii and
bulky crusts \citep{HPY07}. Their cooling can be regulated by
neutrino emission from the crust given by equation~(\ref{e:Lcr1}).

\end{document}